\documentclass[aps,prd,preprint,superscriptaddress,showpacs,floatfix,
preprintnumbers,nofootinbib, a4paper]{revtex4-1}
\usepackage{color}
\usepackage{graphicx}
\usepackage{textcomp}
\usepackage{subfigure}
\usepackage{feynmp}
\usepackage{amsmath,amssymb,array}
\usepackage{enumerate}
\newcommand{\mathsym}[1]{{}}

\newcommand{\ba}{\begin{array}} 
\newcommand{\ea}{\end{array}}
\newcommand{\be}{\begin{equation}}
\newcommand{\ee}{\end{equation}}
\newcommand{\beqa}{\begin{eqnarray}} 
\newcommand{\eeqa}{\end{eqnarray}}

\begin{document} 
\vspace*{1cm}
\title{Clockwork Mechanism for Flavour Hierarchies}
\bigskip
\author{Ketan M. Patel}
\email{ketan@iisermohali.ac.in} 
\affiliation{Indian Institute of Science Education and Research Mohali, Knowledge City, Sector  81, S A S Nagar, Manauli 140306, India \vspace{0.5cm}}

\begin{abstract}
We incorporate clockwork mechanism into the Standard Model flavour sector and show that the observed pattern of fermion masses and mixing can be obtained without any unnaturally small or large parameter in the fundamental theory. By introducing $N_f$ pairs of vectorlike fermions, as clockwork gears, for each generation of the Standard Model fermions and setting up a characteristic clockwork potential, it is shown that the inter-generational mass hierarchies are determined by $N_f$. For a given type of fermions, strong or mild hierarchy in the masses and mixing parameters can be obtained by taking the large or small value of $N_f$. The mechanism is shown to lead to a generalized version of the Froggatt-Nielsen mechanism as an effective description.
\end{abstract} 

\maketitle

\section{Introduction}
\label{intro}
A lack of understanding of the observed hierarchies between the masses of charged fermions and differences between the magnitudes of mixing among the quarks and leptons of the Standard Model (SM) constitutes the so-called flavour puzzle; see for example, a recent review \cite{Feruglio:2015jfa}.  A useful insight into the flavour puzzle can be obtained by representing the various ratios of fermion masses and mixing parameters in terms of the powers of a small parameter, namely the Cabibbo angle, $\lambda = 0.23$. For the masses at the weak scale \cite{Xing:2011aa,Esteban:2016qun}, one obtains 
\beqa \label{mass-ratio}
\frac{m_u}{m_t} &  \approx & \lambda^{8.0}\,,~~\frac{m_d}{m_b}  \approx \lambda^{4.7}\,,~~\frac{m_e}{m_\tau}  \approx \lambda^{5.6}\,,~~\frac{m_{\nu_1}}{m_{\nu_3}}  \approx \lambda^{\infty}-1\,~\nonumber \\
\frac{m_c}{m_t} &  \approx & \lambda^{3.8}\,,~~\frac{m_s}{m_b}  \approx \lambda^{2.7}\,,~~\frac{m_\mu}{m_\tau}  \approx \lambda^{1.9}\,,~~\frac{m_{\nu_2}}{m_{\nu_3}}  \approx \lambda^{1.2}-1\,~ \eeqa
\beqa \label{mass-ratio2}
\frac{m_b}{m_t} &  \approx & \lambda^{2.8}\,,~~\frac{m_\tau}{m_t}  \approx \lambda^{3.1}\, \eeqa
The elements of the quark and lepton mixing matrices denoted by $V$ and $U$, respectively, can also be approximated as \cite{Patrignani:2016xqp,Esteban:2016qun}
\beqa \label{mixing}
|V_{us}| &  \approx & \lambda\,,~~|V_{cb}|   \approx \lambda^{2.2}\,,~~|V_{ub}|  \approx \lambda^{3.8}\,,~\nonumber\\
|U_{e2}| &  \approx & \lambda^{0.4}\,,~~|U_{\mu 3}|   \approx  \lambda^{0.3}\,,~~|U_{e3}|  \approx \lambda^{1.3}~. \eeqa

Although such hierarchies are technically natural, {\it i.e.} they remain almost unaffected under the renormalization group evolution, their existence in the first place still demands an explanation if a complete naturalness is expected from an underlying theory. Several mechanisms have been proposed in which such exponential hierarchies  are generated without requiring very small or large fundamental parameters. For example, it was first pointed out by Froggatt and Nielsen in \cite{Froggatt:1978nt} that the hierarchical structure of quark masses and mixing angles can be induced in an effective theory with a spontaneously broken $U(1)_{\rm FN}$ flavour symmetry. In the simplest realization of the Froggatt-Nielsen (FN) mechanism, a scalar field $\chi$ with $U(1)_{\rm FN}$ charge ${\rm FN}(\chi) = -1$ develops a Vacuum Expectation Value (VEV) which can be conveniently expressed as $\langle \chi \rangle = \lambda / \Lambda$, where $\Lambda$ is the cut-off scale. The different flavours and generations of quarks and leptons possess different charges under the $U(1)_{\rm FN}$, which determine the hierarchies in the masses and mixing parameters. For example, the Yukawa couplings for different types of fermions are obtained in an effective theory as
\be \label{effective-yukawa}
{\cal Y}_u = F_q\, Y_u\, F_{u^c}\,,~~{\cal Y}_d = F_q\, Y_d\, F_{d^c}\,,~~{\cal Y}_e = F_l\, Y_e\, F_{e^c}\,,~~{\cal Y}_\nu = F_l\, Y_\nu\, F_{l}\,,\ee
where $Y_f$ (for $f=u,d,e,\nu$) are $3 \times 3$ matrices with complex elements of ${\cal O}(1)$. We assume that neutrinos are Majorana fermions and $Y_\nu$ represents the coupling matrix of a dimension-five operator which induces small neutrino masses. The matrices $F_f$ are determined from the $U(1)_{\rm FN}$ charges of corresponding fermions and are given by
\be \label{FN-matrices}
F_f = {\rm Diag.}\left( \lambda^{{\rm FN}(f_1)}, \, \lambda^{{\rm FN}(f_2)}, \,  \lambda^{{\rm FN}(f_3) }\right)~, \ee
where $f_i = q_i, u^c_i, d^c_i, l_i, e^c_i$, and FN($f_i$) is the charge of the fermion $f_i$ under $U(1)_{\rm FN}$.

For ${\rm FN}(f_1)  \ge {\rm FN}(f_2) \ge {\rm FN}(f_3) \ge 0$, the mass of $i$th generation of fermion of type $f$ is of the order of $\lambda^{{\rm FN}(f_i) + {\rm FN}(f^c_i)}$, where $f^c_i$ is the charge conjugate partner of $f_i$. Moreover, the mixing between the $i$th and $j$th ($j>i$) generations of fermion of flavour $f$ is proportional to $\lambda^{{\rm FN}(f_i) - {\rm FN}(f_j)}$ \cite{Feruglio:2015jfa}. Following this, the quark mass ratios in Eq. (\ref{mass-ratio}) and the quark mixing parameters in Eq. (\ref{mixing}) are best described by the following choice of FN charges:
\be \label{FN-quark}
{\rm FN}(q_i) = {\rm FN}(u^c_i) = \left\{ 4,2,0\right\}\,,~~{\rm FN}(d^c_i) = \left\{s+1, s, s\right\},~(s=0,1,2,3)~. \ee
Here, $s \approx 3$ reproduces the correct ratio $m_b/m_t$  (as given in Eq. (\ref{mass-ratio2})) in the effective theories with only one Higgs doublet, like the SM. In the theories like the two-Higgs-doublet models, the ratio $m_b/m_t$ depends also on a ratio of VEVs of the two Higgs, namely, $v_u/v_d \equiv \tan\beta$.  Depending on the value of $\tan \beta$, one can choose any value for $s$ between 0 and 3 in order to reproduce the correct $m_b/m_t$. The FN mechanism can also be extended to the lepton sector with the following choice of FN charges for the leptons:
\be \label{FN-lepton}
{\rm FN}(l_i) = \left\{s+1,s,s\right\}\,,~~{\rm FN}(e^c_i) = \left\{ 4,2,0\right\}~.
\ee
A possibility of no or mild hierarchy in the neutrino masses also allows a choice ${\rm FN}(l_1) ={\rm FN}(l_2) = {\rm FN}(l_3)$. However, ${\rm FN}(l_1) > {\rm FN}(l_2)$ helps in obtaining small $|U_{e3}|$. The choice ${\rm FN}(l_2) = {\rm FN}(l_3)$ is favoured by the large atmospheric mixing angle.

In this article we propose a new mechanism to obtain an effective theory with FN-like structure. The mechanism is based on recently proposed Clockwork (CW) phenomena \cite{Giudice:2016yja} (see also \cite{Choi:2014rja,Choi:2015fiu,Kaplan:2015fuy} for earlier versions) in which exponentially suppressed interactions are induced in a theory containing all the fundamental couplings of natural magnitude. An underlying theory includes multiple fields, namely the clockwork gears, and a potential containing interactions between only the adjacent CW gears. When an interaction is introduced between an external sector and the field at one of the end sites of the CW chain, the arrangement generates interaction between the external sector and the field at the other end with exponentially enhanced or suppressed coupling. The CW mechanism can be applied to the various kinds of fields \cite{Giudice:2016yja} and it has already been studied for its phenomenological applications for the Higgs sector \cite{Giudice:2016yja,Kim:2017mtc}, inflation \cite{Choi:2014rja,Kehagias:2016kzt,Im:2017eju}, axion \cite{Choi:2015fiu,Kaplan:2015fuy,Coy:2017yex}, dark matter \cite{Hambye:2016qkf}, neutrino masses \cite{Hambye:2016qkf,Park:2017yrn,Ibarra:2017tju} and other sectors \cite{Ahmed:2016viu,Kehagias:2017grx,Hong:2017tel,Lee:2017fin,Antoniadis:2017wyh,Ibanez:2017vfl,Ben-Dayan:2017rvr}. We explore this mechanism in the context of the SM flavour puzzle here. A preliminary version of  fermionic CW is already discussed in \cite{Giudice:2016yja}. We show that it can  straightforwardly be generalized to the SM flavours such that one can obtain an effective FN-like structure. The number of CW gears plays a role of FN charges in this mechanism and by taking a different number of gears for different generations, one can explain intergenerational mass hierarchies.

We propose a more attractive extension of the above mechanism, namely Flavoured Clockworks (FCW). Unlike the previous case, the different generations of a given type of fermion have the same number of CW gears in this mechanism. They are allowed to couple with each other, maintaining the nearest-neighbour interaction structure of the CW mechanism. It is shown that, in this case the effective FN-like suppression factors arise as the eigenvalues of a matrix which is proportional to $Q^N$, where $Q$ is an $n_f \times n_f$ matrix for $n_f$ number of families such that the magnitude of elements of $Q$ is less than unity while $N$ being the number of CW gears. Even if there is no large hierarchy between the eigenvalues of $Q$, the eigenvalues of $Q^N$ can be hierarchical. The number of CW gears therefore control the hierarchy and one can obtain strong or mild inter-generational hierarchies for different types of fermion through an appropriate choice of $N$. The FCW mechanism provides a generalized ultraviolet completion of the FN mechanism such that an effective theory can generate fractional FN charges. Moreover the effective FN charges of different generations are determined by a parameter $N$.

We briefly review the CW mechanism and its simplest extension to the SM fermions in the next section. In Section \ref{FCW}, we present FCW mechanism and discuss its features. The viability of FCW in reproducing the known structure of the masses and mixing of SM fermions is demonstrated in Section \ref{pheno}. The study is summarized in Section \ref{summary}.

\section{Clockwork Mechanism}
\label{mechanism}
The CW mechanism for fermions includes a number of fields, termed as the clockwork gears, and a chiral  symmetry which keeps one of these gears massless \cite{Giudice:2016yja}. Consider $N$ left-handed Weyl fermions $f_a$ ($a=1,...,N$) and their $N+1$ partners, $f^c_b$ ($b=0,1,...,N$), with opposite gauge quantum numbers. Let us consider a global symmetry which is a product of several $U(1)$ factors: $G = \prod_{a,b} U(1)_{L,a} \times U(1)_{R,b}$. Under the $G$, the fields $f_a$ and $f^c_b$ have charges $(1,0)$ and $(0,1)$, respectively. The symmetry $G$ is then broken by $N$ mass terms giving rise to $N$ massive Dirac fermions and leaving one linear combination of $f^c_b$ as a massless fermion. The massless state is a result of a subgroup of $G$, namely $U(1)_{R,0}$, which remains unbroken by the $N$ mass terms. The Lagrangian governing the CW mechanism is given by 
\be \label{CW-toy}
{\cal L} = i \bar{f}^c_0 \gamma^\mu \partial_\mu f^c_0\, + \sum_{a=1}^{N} \left( i \bar{f}_a \gamma^\mu \partial_\mu f_a\, +\, i \bar{f}^c_a \gamma^\mu \partial_\mu f^c_a\, -  \left( M\, f_a f^c_a - m\, f_a f^c_{a-1} + {\rm h.c.} \right) \right)~,\ee
where $m$ and $M$ are the mass parameters and for simplicity, they are assumed to be universal for all $a$. The $m$ and $M$ can be treated as spurions with charges $(-1,-1)$ under  $U(1)_{L,a} \times U(1)_{R,b-1}$ and $U(1)_{L,a} \times U(1)_{R,b}$, respectively. We assume $m<M$ and both of them are taken as real parameters for simplicity. The fermions at the $b^{\rm th}$ site, when integrated out by solving Euler-Lagrange equations of motions for $f_b$ and $f^c_b$, one obtains
\be \label{}
f_b = 0\,,~~~f^c_{b} = \left(\frac{m}{M}\right) f^c_{b-1} \equiv q\, f^c_{b-1}\, ~~~(b=1,2,...,N)~.
\ee
Interestingly, $f^c_{b-1}$ has an overlap with the fermion located at the $b^{\rm th}$ site suppressed by a factor $q$.  After integrating out all the massive $N$ pairs of $f_a$ and $f_a^c$ from Eq. (\ref{CW-toy}), one obtains an effective Lagrangian 
\be \label{}
{\cal L}_{\rm eff} =  i \bar{f}^c_0 z \gamma^\mu \partial_\mu f^c_0\,  \equiv i \bar{f}^c  \gamma^\mu \partial_\mu f^c\,, \ee
where
\be \label{}
z = \sum_{b=0}^N q^{2b} = \frac{1-q^{2(N+1)}}{1-q^2}~~~{\rm and}~~~f^c = \sqrt{z}f^c_0 = \sqrt{z}\, q^{-N} f^c_N~.
\ee
Here $f^c$ is a massless fermion with canonically normalized kinetic term. It has an overlap with the fermion at the $N$th site which is suppressed by $q^N$. This is an essence of the CW mechanism which can be utilized to generate small couplings in an effective theory starting with natural couplings in the fundamental theory.

If the Yukawa interaction between $f^c_b$ and an another field $\psi$ is introduced at the $N$th site in the fundamental theory with a coupling of natural size then
\be \label{}
{\cal L}_Y = -  y\, \psi H  f^c_N +{\rm h.c.} = - \frac{y\, q^N}{\sqrt{z}} \psi\, H\, f^c +{\rm h.c.}~.
\ee
One obtains small Yukawa coupling in an effective theory which is suppressed by a factor $q^N$. 

This mechanism can straightforwardly be applied to all the SM fermions. In the simplest case, one can introduce $N_{f_i}$ clockwork gears for a given generation of the SM fermion of type $f$. With a universal value of $q \approx \lambda$ and $N_{f_i} = {\rm FN}(f_i)$, the CW mechanism leads to an effective theory equivalent to the one obtained using FN mechanism. The $N_{f_i}$ for different $f_i$ can directly be chosen from Eqs. (\ref{FN-quark}) and (\ref{FN-lepton}) such that they reproduce realistic spectrum of fermion masses and mixing. In this spirit, the CW theory can be seen as an ultraviolet description of the FN mechanism. Note that in this simplest case, the CW gears of different generations in a given fermion sector are not allowed to couple with each other in order to sustain the inter-generational hierarchies created by different number of gears for each generation.  Although such an arrangement can be justified by enhancing the symmetry of fundamental theory, we show that it is not necessarily needed.

\section{Flavoured Clockworks}
\label{FCW}
We now generalize the CW mechanism to FCW. Unlike the mechanism discussed in the previous section, the CW gears of different generations are allowed to interact with each other, maintaining the nearest neighbour interaction structure of the CW mechanism. For $n_f$ generations of a given type of fermion, the FCW gears consist of the fields: $f^{(i)}_a$ ($a=1,2,...,N$; $i=1,...,n_f$) and $f^{c(j)}_b$ ($b=0,1,...,N$; $j=1,...,n_f$).  The Lagrangian in Eq. (\ref{CW-toy}) can then be generalized to accommodate FCW as the following:
\beqa \label{L-FCW}
{\cal L} &=& \sum_{i,j=1}^{n_f} \left(i \bar{f}^{c(i)}_0 \gamma^\mu \partial_\mu f^{c(i)}_0\, + \sum_{a=1}^{N} \left( i \bar{f}^{(i)}_a \gamma^\mu \partial_\mu f^{(i)}_a\, +\, i \bar{f}^{c(i)}_a \gamma^\mu \partial_\mu f^{c(i)}_a\, \right. \right. \nonumber \\
& - &  \left. \left. \left( M_{ij}\, f^{(i)}_a f^{c(j)}_a - m_{ij}\, f^{(i)}_a f^{c(j)}_{a-1} + {\rm h.c.} \right) \right) \right)~,\eeqa
where $M$ and $m$ are now $n_f \times n_f$ matrices in generation space. For the vanishing $M$ and $m$, the above Lagrangian possesses a global $\prod_{a,b} U(n_f)_{L,a} \times U(n_f)_{R,b}$ symmetry under which $n_f$ generations of $f_a$ and $f^c_b$ transform as $(n_f,1)$ and $(1,n_f)$, respectively. This symmetry is broken by the spurions $M$ and $m$, which transform as $(\overline{n}_f,\overline{n}_f)$ under $U(n_f)_{L,a} \times U(n_f)_{R,b}$ and $U(n_f)_{L,a} \times U(n_f)_{R,b-1}$, respectively. For simplicity, we choose the matrices $m$ and $M$ to be universal for different gears. Integrating out the $f_b^{(j)}$ and $f^{c(j)}_{b}$ from Eq. (\ref{L-FCW}) leads to the following solutions,
\be \label{}
f^{(j)}_b = 0\,,~~~f^{c(j)}_{b} = \sum_{k=1}^{n_f} Q_{jk} f^{c(k)}_{b-1}\, ~~~(b=1,2,...,N),
\ee
where  $Q_{ij} = (M^{-1} m)_{ij}$. We further assume both $m$ and $M$ to be real and $|Q_{ij}| < 1$.

After integrating out all the $n_f \times N$ pairs of gears, one obtains an effective Lagrangian
\be \label{eff}
{\cal L}_{\rm eff} =  i \bar{f}^c_0 Z \gamma^\mu \partial_\mu f^c_0\,, \ee
for $n_f$ massless fields $f^{c}_0 = (f^{c(1)}_{0},...,f^{c(n_f)}_{0})^T$, where
\be \label{}
Z = {\bf 1} + Q^\dagger Q + Q^{\dagger 2} Q^2 + ... +Q^{\dagger N} Q^N~,
\ee
and $Q$ is an $n_f \times n_f$ matrix which can be diagonalized using biunitary transformations. This leads to
\beqa \label{}
V_Q\,Q\, U_Q^\dagger\, & = & {\rm Diag.}(q_1,...,q_{n_f}) \equiv \tilde{Q} \nonumber \\
U_Q\, Z\, U_Q^\dagger\, & = & {\rm Diag.}(z_1,...,z_{n_f}) \equiv \tilde{Z}~. \eeqa
The effective Lagrangian in Eq. (\ref{eff}) can be rewritten as
\be \label{}
{\cal L}_{\rm eff} =  i \bar{f}^c  \gamma^\mu \partial_\mu f^c\,,\ee
with $f^c = \sqrt{\tilde{Z}}\, U_Q\, f^c_0$ being $n_f$ massless modes with a canonically normalized kinetic term.

A Yukawa interaction defined at the $N$th site then leads to 
\be \label{}
{\cal L}_Y = -\, \psi\, Y\,  H  f^c_N +{\rm h.c.} = -\, {\psi}\, \left(Y V_Q^\dagger\, \tilde{Q}^N  \sqrt{\tilde{Z}^{-1}}\right)\,   H\, f^c +{\rm h.c.},
\ee
where $\psi = (\psi_1,...,\psi_{n_f})^T$ are $n_f$ number of fermions and $H$ is a scalar. The  $n_f \times n_f$ Yukawa coupling matrix $Y$ (and therefore $YV_Q^\dagger$) has elements of magnitude of ${\cal O}(1)$. The diagonal matrix $\tilde{Q}^{N} \sqrt{\tilde{Z}^{-1}}$ acts as a suppression factor and generates a hierarchical structure in the effective Yukawa couplings.

The FCW mechanism described above can be applied to the different SM fermion flavours: $f_i = q_i, u^c_i, d^c_i, l_i, e^c_i$. Generalizing the above mechanism to such fermions, we define
\be \label{FCW-F}
\tilde{F}_f \equiv  \tilde{Q}_f^{N_f} \sqrt{\tilde{Z_f}^{-1}} = {\rm Diag.}\left( \lambda^{n(f_1)}\,, ..., \lambda^{n(f_{n_f})} \right)~,\ee
where $Q_f$ and $N_f$ are the coupling matrix and number of gears for a given SM fermion sector, respectively. The elements of $\tilde{F}_f$ are expressed as the powers of $\lambda$ in order to compare the FCW framework with the FN mechanism. The $\tilde{F}_f$ when compared with $F_f$ in Eq. (\ref{FN-matrices}), the powers $n(f_i)$ can be seen as an effective FN charge of fermion $f_i$. Unlike in the FN mechanism, the effective charge can be a fractional number in this case. The major difference between the FCW and simple CW discussed in the previous section is that the flavour hierarchy in the FCW mechanism arise from multiple products of the matrix $Q_f$. Even if the elements of $Q_f$ are chosen randomly such that $|(Q_f)_{ij}| < 1$, the eigenvalue spectrum of $Q_f^{N_f}$ turns out to be hierarchical. This follows from a more general observation made recently in \cite{vonGersdorff:2017iym}. It was observed that products of matrices with elements of random ${\cal O}(1)$ numbers has hierarchical eigenvalue spectrum. If the matrix $Q_f$ has eigenvalues $\lambda_i$, ($|\lambda_i| < 1$ and $i=1,...,n_f$) then the eigenvalues of $Q_f^{N_f}$ are given by $\lambda_i^{N_f}$. Therefore even a mild hierarchy between the $\lambda_i$'s becomes strong in $\lambda_i^{N_f}$'s when the number of CW gears are large. It can be seen from Eq. (\ref{FCW-F}) that the value of $N_f$ almost determines the effective FN charges for all the generations of a given fermion flavour. This is the main feature of the FCW mechanism which distinguishes it from a simple extension of the CW mechanism to the SM flavour sector discussed in the previous section.

The effective theory of the flavour obtained from the FCW resembles with the ones obtained from the theories with extra spatial dimensions upon compactification \cite{ArkaniHamed:1999dc,Kaplan:2001ga,Grossman:1999ra,Gherghetta:2000qt}. In the latter, the exponential hierarchy in the Yukawa couplings in four dimensions is produced through different localization of fermion fields in the extra dimension. The four-dimensional fermion fields convoluted with the profile factor in extra dimension produces strong or weak coupling in the effective theory depending upon their localization. In the FCW, the  gears mimic the role of the extra dimension and the number of CW gears determines the hierarchy among the effective couplings of the theory. Both the theories can lead to effective FN description and can accommodate fractional FN charges for fermions. The details of similarities between the four-dimensional CW mechanism and the theories with extra dimension have been discussed in \cite{Giudice:2016yja,Craig:2017cda,Giudice:2017suc}. It is found that they are not exactly the same \cite{Craig:2017cda}.

We also note that the FCW framework described in the above can be seen as a specific case of the recently proposed mechanism \cite{vonGersdorff:2017iym} which is based on an observation that products of random ${\cal O}(1)$ matrices possesses hierarchical spectrum. In Eq. (\ref{FCW-F}), the suppression factor arises from a multiple product of the $Q_f$ matrix which is made up of elements of ${\cal O}(1)$. One can also work with nonuniversal ({\it i.e.} gear-dependent) coupling matrices which provide a characteristic example of the mechanism proposed in \cite{vonGersdorff:2017iym}.

\section{Fermion mass pattern from Flavoured Clockworks}
\label{pheno}
\begin{figure}[!ht]
\centering
\subfigure{\includegraphics[width=0.49\textwidth]{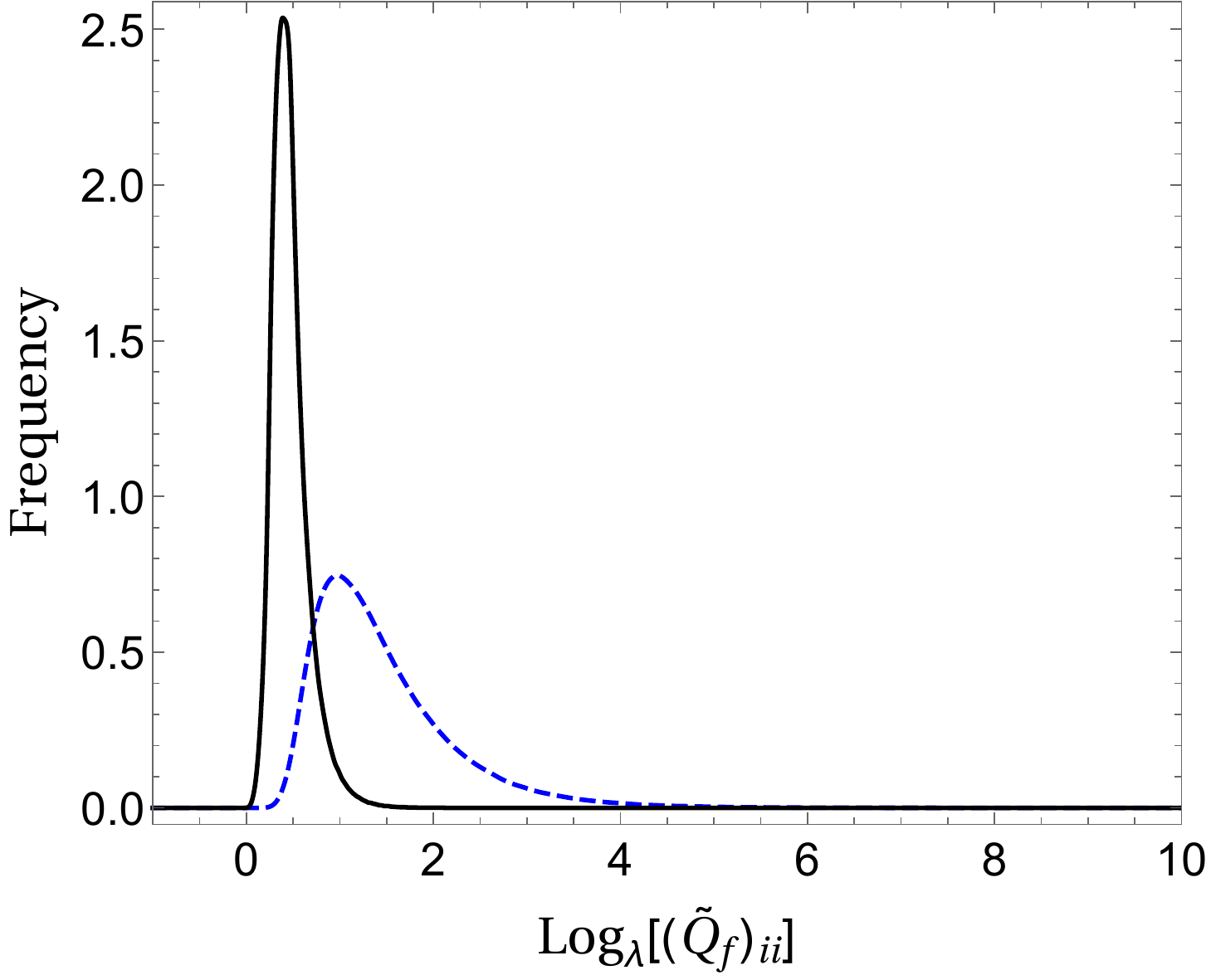}}
\subfigure{\includegraphics[width=0.49\textwidth]{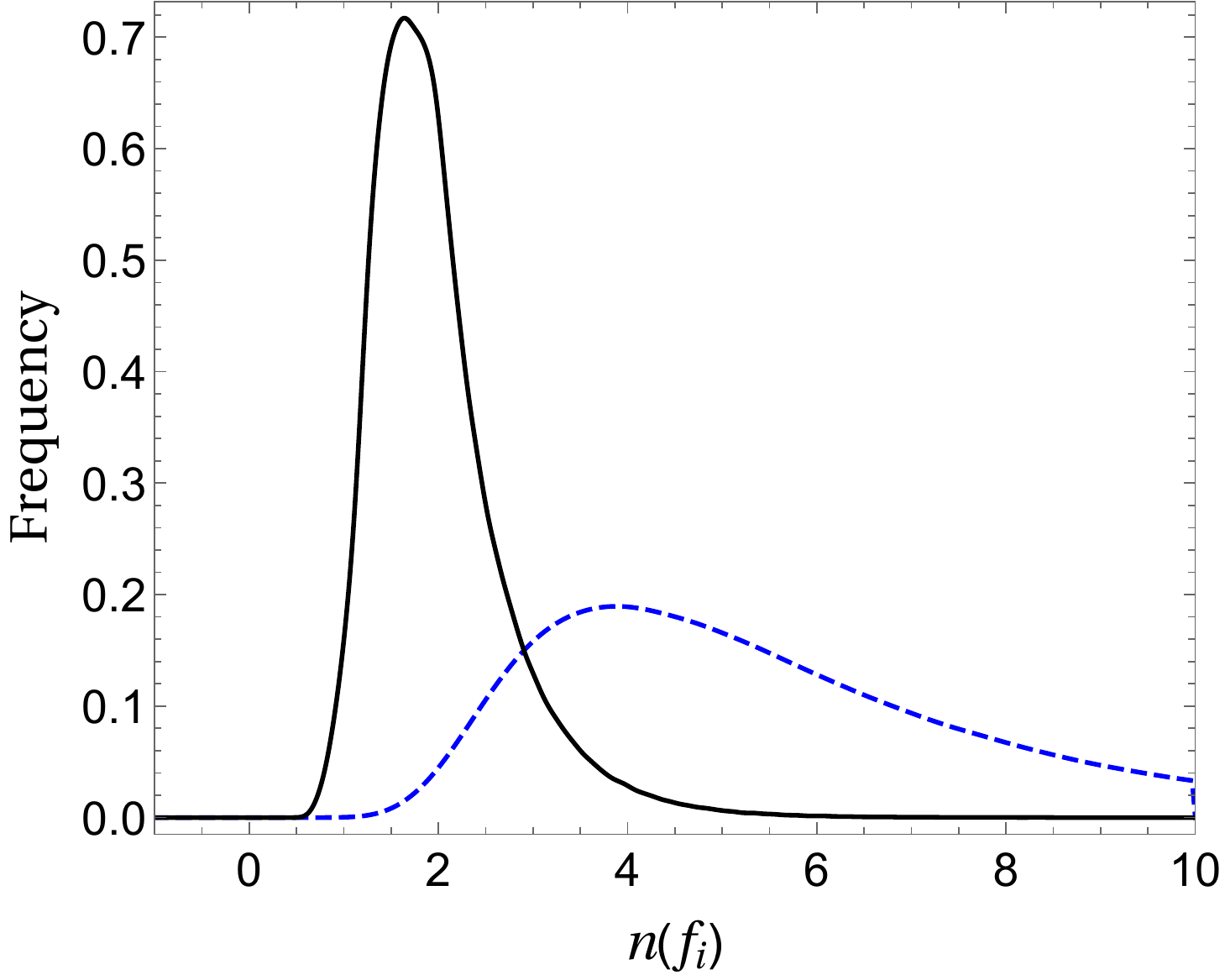}}\\
\caption{Left: The eigenvalue spectrum of the $2\times 2$ real matrix $Q_f$, elements of which are randomly chosen form a uniform distribution of numbers between $-0.5$ and $0.5$. Right: The distributions of  $n(f_2)$ (solid black line) and  $n(f_1)$ (dashed blue line) obtained from Eq. (\ref{FCW-F}) for such $Q_f$ and with $N_f=4$.}
\label{fig1}
\end{figure}

The result obtained in Eq. (\ref{FCW-F}) can be used to simulate the effective FN-like charges for various fermions using appropriate coupling matrix $Q_f$ and selecting the appropriate number of CW gears $N_f$. For example, in case of  $f=q,u^c,e^c$, the Yukawa couplings for the third generation are required to be as large as ${\cal O}(1)$. Therefore, one can introduce FCW only for the first two generations, {\it i.e.} with $n_f = 2$. Taking $Q_f$ as $2 \times 2$ matrix whose elements are chosen from a random uniform distribution of numbers within the range $[-0.5, 0.5]$, one obtains the eigenvalue spectrum as shown in the left panel of Fig. \ref{fig1}. 
\begin{figure}[!ht]
\centering
\subfigure{\includegraphics[width=0.49\textwidth]{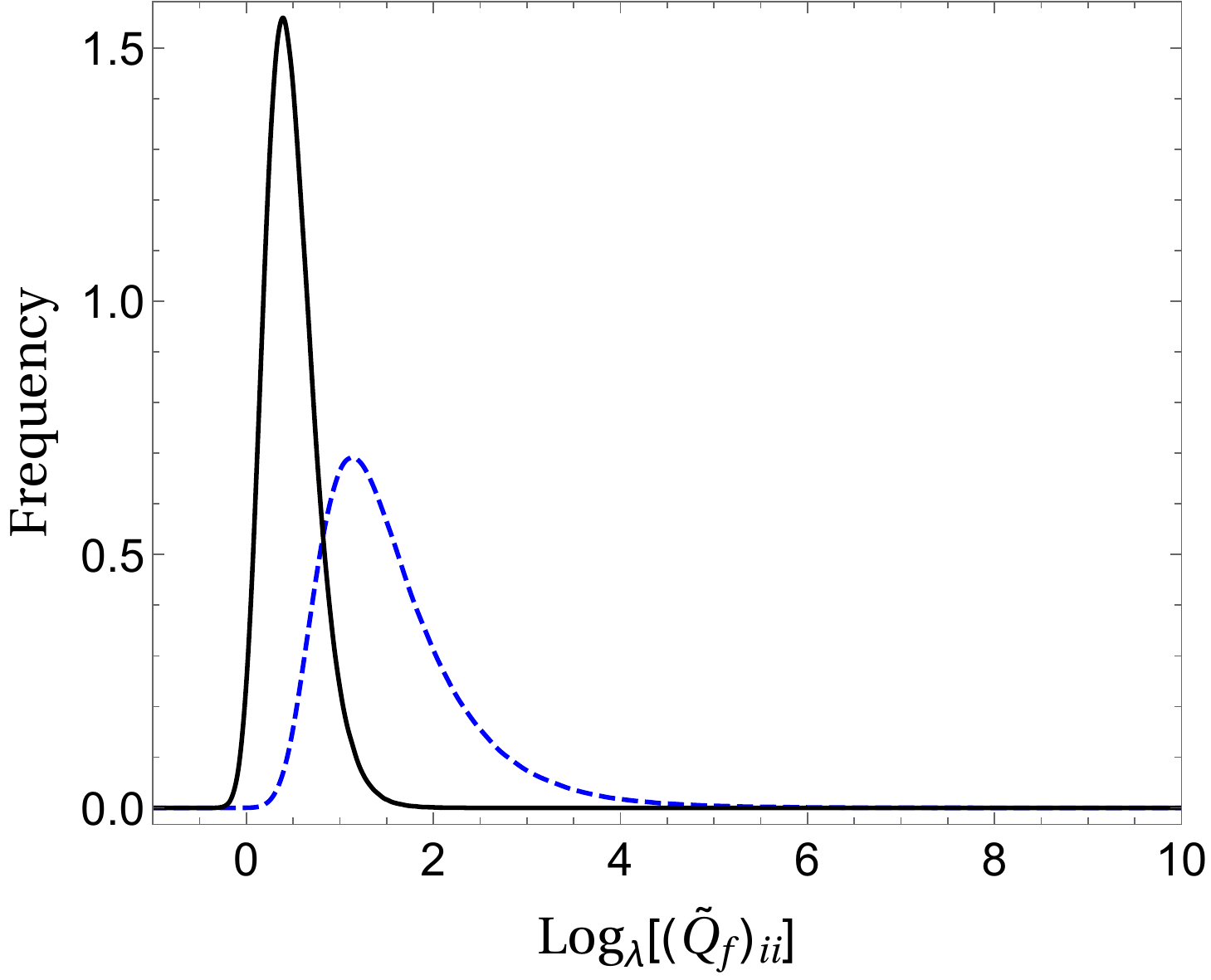}}
\subfigure{\includegraphics[width=0.49\textwidth]{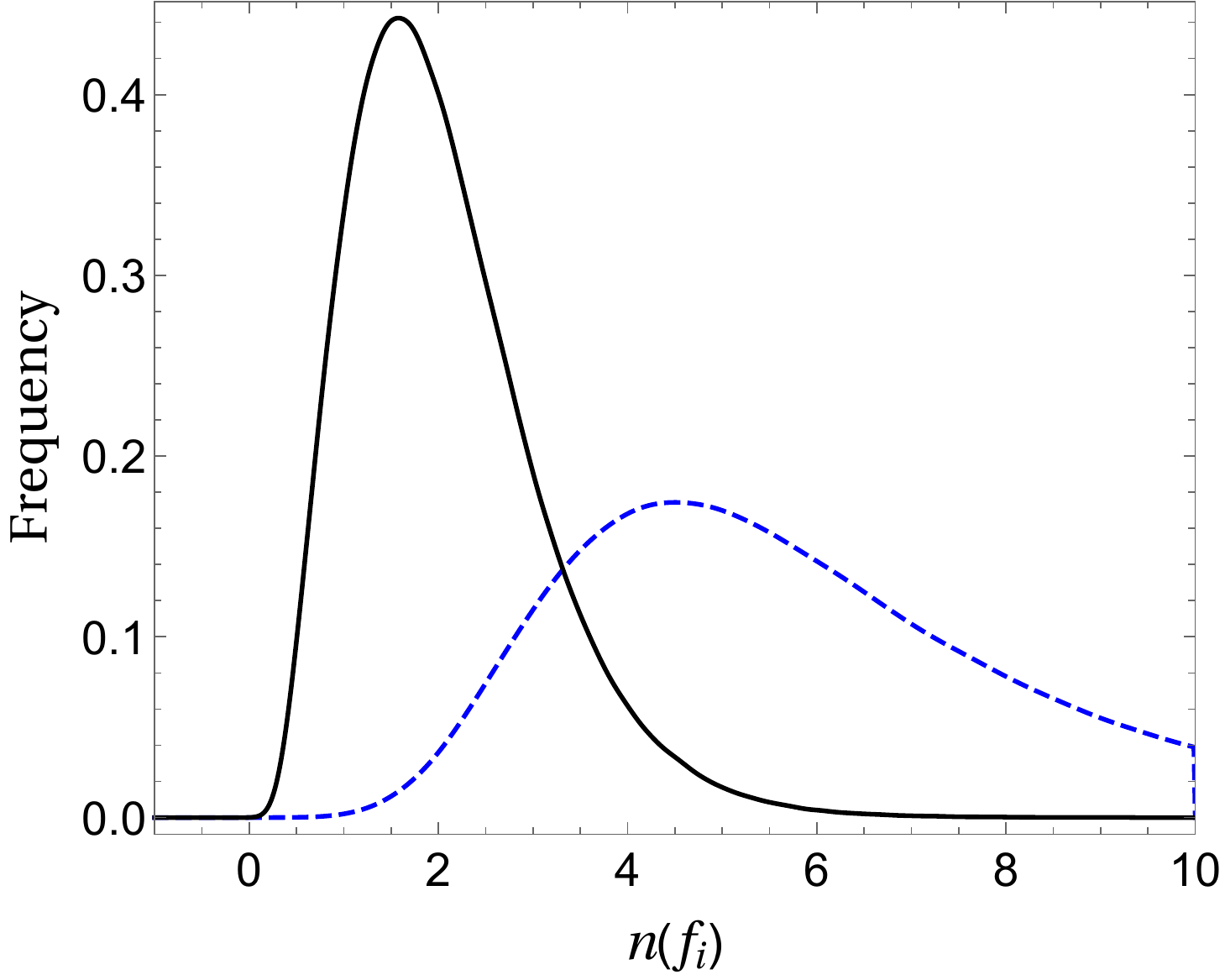}}\\
\caption{Same as Fig. \ref{fig1} for the elements of $Q_f$ randomly chosen form normal distribution with $\mu=0$ and $\sigma = 0.3$.}
\label{fig2}
\end{figure}
The distributions of the eigenvalues $\lambda_{1,2}$ peak at the values $\lambda_1 \approx \lambda$ and $\lambda_2 \approx  \lambda^{0.5}$. The corresponding distributions of the eigenvalues of $\tilde{F}_f $ obtained using Eq. (\ref{FCW-F}) and $N_f =4$ are displayed in the right panel of Fig. \ref{fig1}. The effective FN-like charges takes the values $n(f_1) \approx 4$ and $n(f_2) \approx 2$.  As it is expected, the FCW scales the distribution $\lambda_i$ to $\lambda_i^{N_f}$ and one gets the exponential hierarchy in the effective Yukawa couplings. We also repeat the analysis for $Q_f$ whose elements are randomly chosen from a normal distribution centered at $\mu=0$ with standard deviation $\sigma=0.3$. The corresponding results are displayed in Fig. \ref{fig2}. These choices of $Q_f$ and $N_f=4$ generate viable effective FN charges for $f=q,u^c,e^c$ as can be seen from Figs. \ref{fig1}, \ref{fig2} and Eqs. (\ref{FN-quark}, \ref{FN-lepton}).

\begin{figure}[!ht]
\centering
\subfigure{\includegraphics[width=0.49\textwidth]{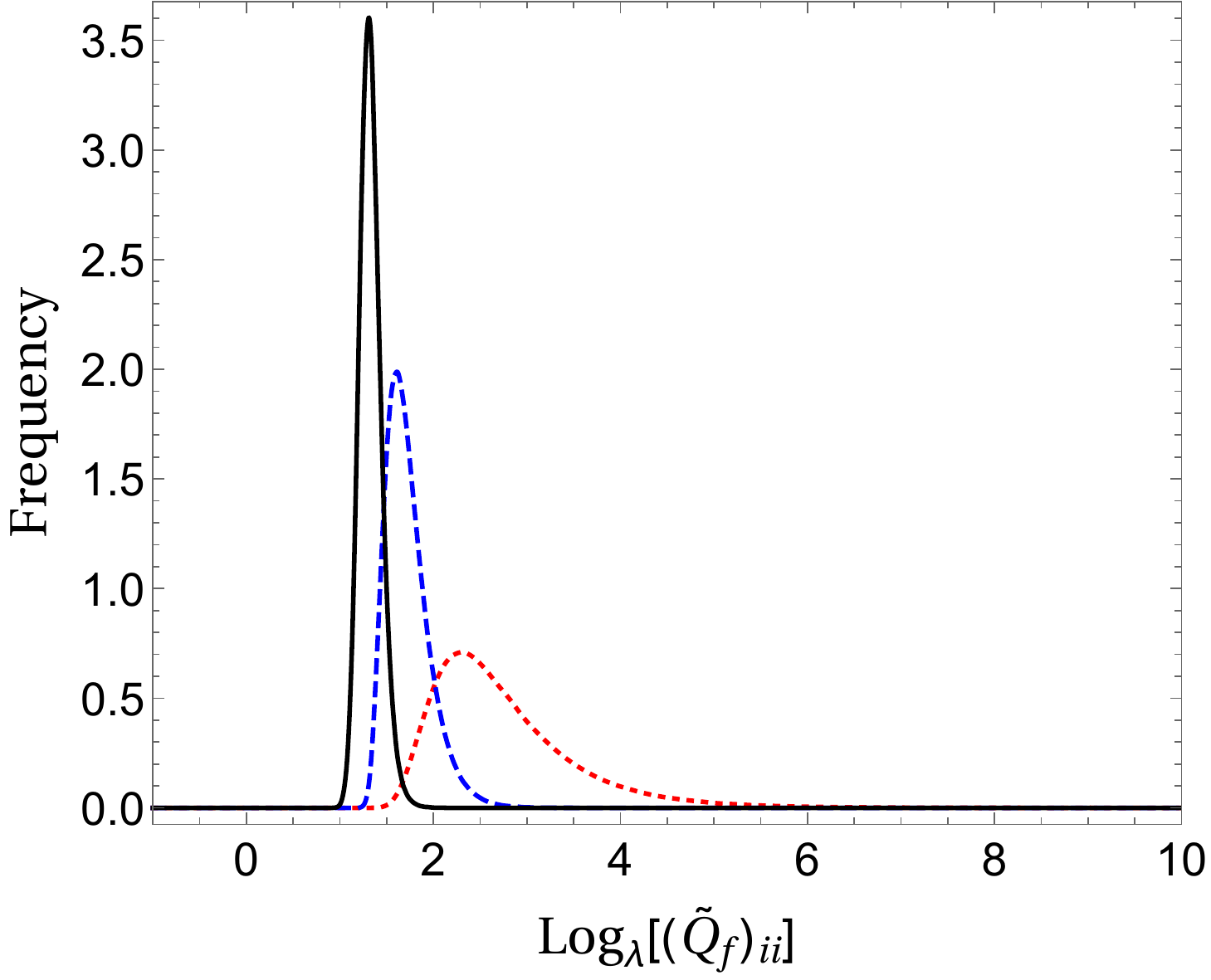}}
\subfigure{\includegraphics[width=0.49\textwidth]{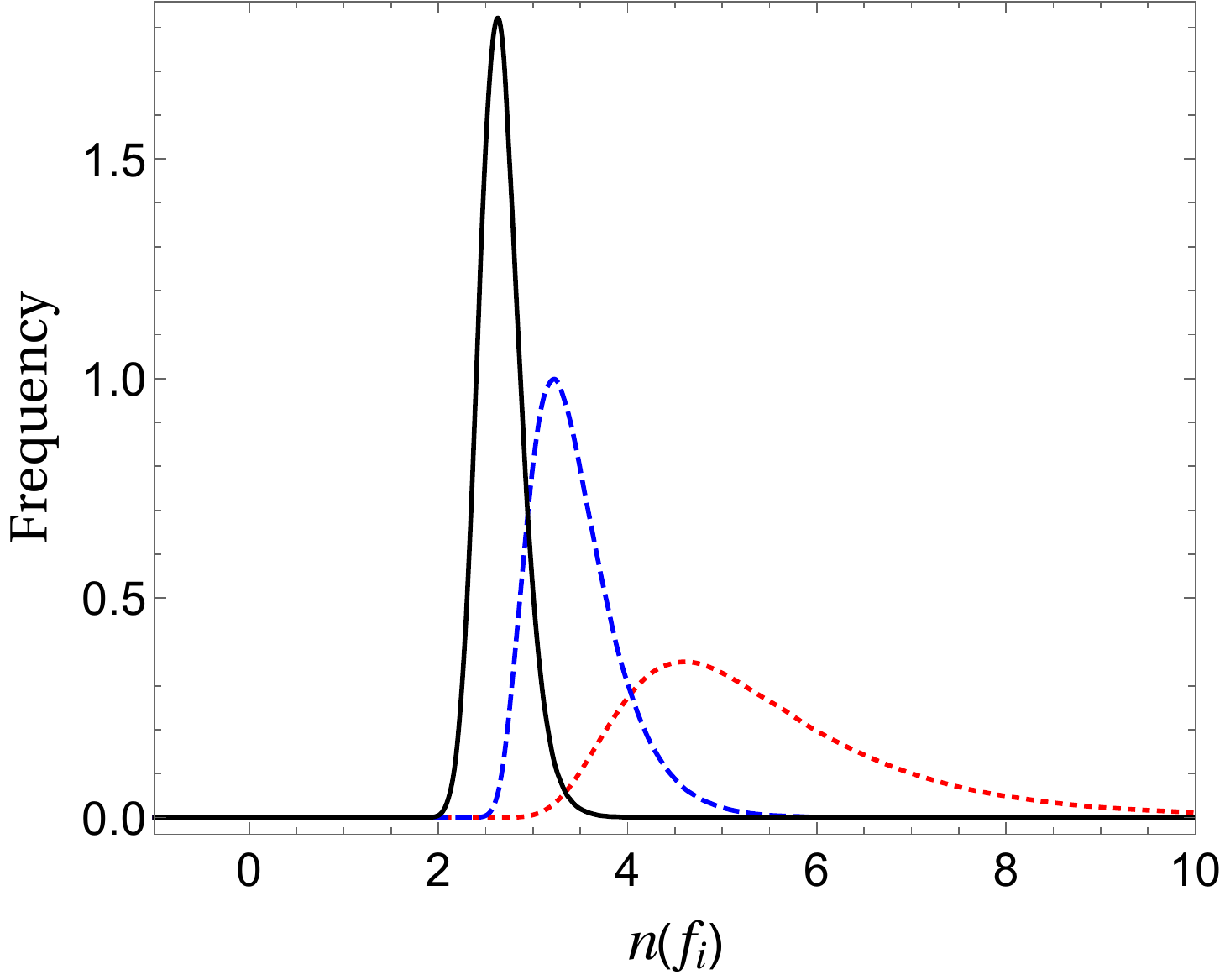}}\\
\caption{Left: The eigenvalue spectrum of $3\times 3$ real matrix $Q_f$, elements of which are randomly chosen form a uniform distribution of numbers between $-0.1$ and $0.1$. Right: The distributions of $n(f_3)$ (solid black line),  $n(f_2)$ (dashed blue line) and $n(f_1)$ (dotted red line) obtained from Eq. (\ref{FCW-F}) for such $Q_f$ and with $N_f=2$.}
\label{fig3}
\end{figure}

The effective FN-like charges for the down-type quarks and charged leptons can be obtained in a similar way. For $s=3$ in Eqs. (\ref{FN-quark}, \ref{FN-lepton}), one can introduce FCW for all the three generations. In this case we choose the elements of $3\times 3$ matrix $Q_f$ from a uniformly distributed random number in $[-0.1,0.1]$ and introduce $N_f = 2$ gears. The distributions of the eigenvalues of $Q_f$ and $\tilde{F}_f $ obtained from Eq. (\ref{FCW-F}) are displayed in Fig. \ref{fig3}.
\begin{figure}[!ht]
\centering
\subfigure{\includegraphics[width=0.49\textwidth]{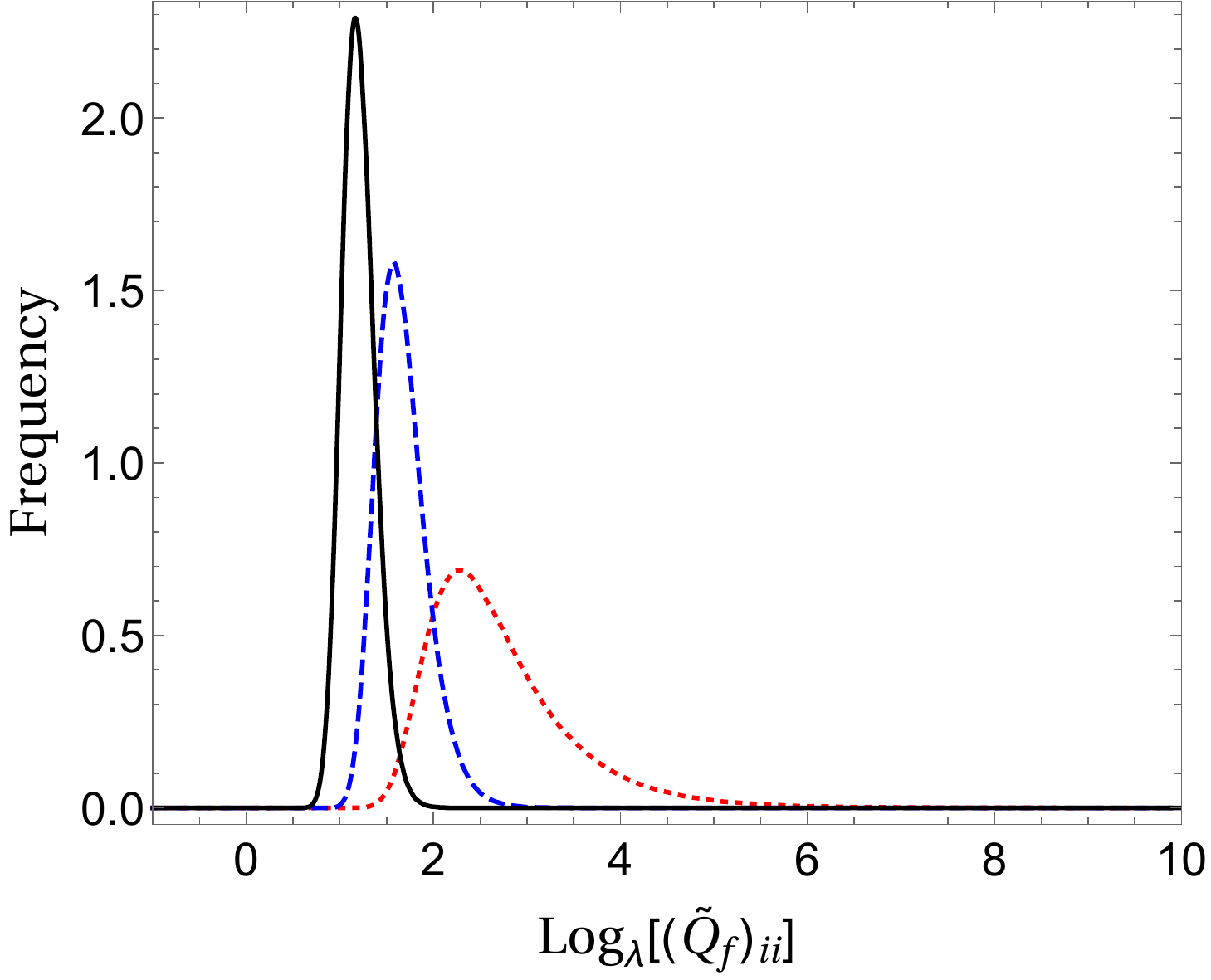}}
\subfigure{\includegraphics[width=0.49\textwidth]{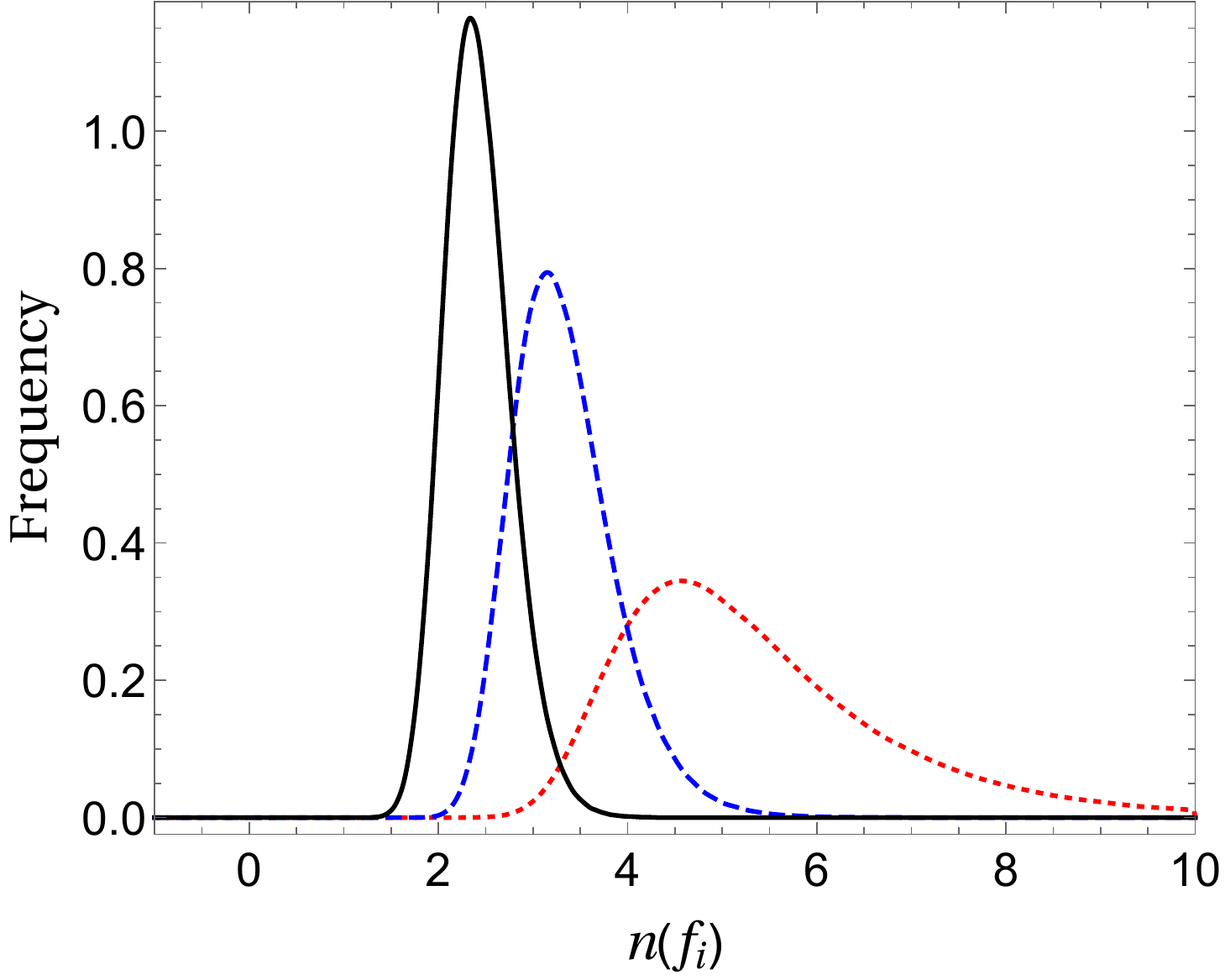}}\\
\caption{Same as Fig. \ref{fig3} for elements of $Q_f$ randomly chosen form a normal distribution with $\mu=0$ and $\sigma = 0.07$. }
\label{fig4}
\end{figure}
A similar analysis is carried out in which elements of $Q_f$ are chosen from a normal distribution centered at $\mu=0$ with standard deviation $\sigma = 0.07$ and its results are displayed in Fig. \ref{fig4}. Note that the largest eigenvalue of $Q_f$ is of ${\cal O}(\lambda)$, which is because of the relatively smaller values chosen for the elements of $Q_f$. A mild hierarchy in the spectrum of $n(f_i)$ is due to the small number of CW gears. The obtained distributions of  $n(f_i)$ displayed in Figs. \ref{fig3} and \ref{fig4} are in good agreement with the desired FN charges for $f=d^c,l$ as can be seen from Eqs. (\ref{FN-quark}, \ref{FN-lepton}) with $s=3$. 

The FN charges of $d^c_i$ and $l_i$ corresponding to $s<3$ in Eqs. (\ref{FN-quark}, \ref{FN-lepton}) can be obtained by taking appropriate range for $Q_f$. For example, $(Q_f)_{ij} \in [-0.2, 0.2]$ favours $s=2$ while $(Q_f)_{ij} \in [-0.3, 0.3]$ is found in good agreement with $s=1$. The choice $s=0$, however, can be best described by two flavour FCW similar to the case for $f=q,u^c,e^c$ described earlier. In all these cases, the number of CW gears remains $N_f=2$, which leads to relatively mild hierarchy in the masses of charged leptons and down-type quarks. $N_f=2$ for $f=l_i$ and $N_f=4$ for $f=q_i$ provide an explanation for large leptonic mixing angles and hierarchical and small quark mixing angles, respectively. The differences in the ranges of $Q_f$ for $f=d^c,l$ and $f=q,u^c,e^c$ can be justified by using different values for the VEVs of spurions $m$ and/or $M$ in Eq. (\ref{L-FCW}).

It is further possible to further improve upon the resulting distributions of $n(f_i)$ by guiding the interactions between the CW gears. For example, it can be noticed form Fig. \ref{fig2} that the distributions of $n(f_2)$ and $n(f_3)$ are peaking at slightly different values. As described earlier, a perfect alignment  $n(f_2) = n(f_3)$ helps in obtaining large or maximal atmospheric mixing angle. This can be achieved if the spurions $M$ and $m$ in Eq. (\ref{L-FCW}) leave an unbroken $O(2)_{R,b}$ subgroup of the full symmetry group of CW theory. The resulting $Q_f$ possesses two identical eigenvalues, let us say $\lambda_2 = \lambda_3$ and leads to $n(f_2) = n(f_3)$ in the effective theory. Therefore, the maximality of atmospheric mixing angle can still be explained in a theory with anarchic Yukawa couplings if an appropriate symmetry governs the interaction among the CW gears.

The FCW presented in the above can straightforwardly be implemented in $SU(5)$ grand unified theories. The SM fermions of a given generation are accommodated in ${\bf 10} = \{ q,u^c,e^c \}$ and $\overline{\bf 5} = \{l,d^c \}$ representations of $SU(5)$. In the above we have found that $Q_q = Q_{u^c} = Q_{e^c} \equiv Q_{{\bf 10}}$,  $Q_{l}=Q_{d^c} \equiv Q_{\overline{\bf 5}}$, $N_q=N_{u^c}=N_{e^c} \equiv N_{\bf 10} = 4$, and $N_{l}=N_{d^c} \equiv N_{\overline{\bf 5}} =2$ which are consistent with $SU(5)$ unification. An implementation of the FCW mechanism within $SO(10)$-based grand unified theories is however a challenging task. It is known that the FN mechanism in its simplest form cannot be implemented in $SO(10)$ theories due to complete unification of all the fermions of a given generation in the ${\bf 16}$-plet of $SO(10)$. However there exist frameworks in which an effective FN-like flavour structure consistent with the flavour spectrum is obtained from $SO(10)$ theory in higher dimension \cite{Kitano:2003cn,Feruglio:2014jla,Feruglio:2015iua}. Considering the resemblance between the FCW and higher-dimensional theories, it would be interesting to explore a possibility to implement FCW in four-dimensional $SO(10)$ grand unified theories.

Finally, we comment on the mass scale of CW gears. The fermion masses are technically natural as they are protected by chiral symmetry. The hierarchies generated at any scale remain unaltered under the renormalization group evolution. Therefore, the scale of vectorlike fermions is not constrained by any naturalness criteria. In other words, the mechanism crucially depends on the ratio, $Q=M^{-1}m$ in Eq. (\ref{L-FCW}), and not on the size of individual $m$ or $M$. There exists a lower bound on the masses of such fermions because of their absence in the direct and indirect searches carried out so far. However, the masses of such fermions can take any value between a few TeV and the Planck scale.

\section{Summary}
\label{summary}
We have implemented the recently proposed CW mechanism into the SM flavour sector. It is shown that one can obtain the exponential hierarchies in the masses of fermions without introducing any unnaturally small or large parameter in the fundamental theory. This is achieved by incorporating $N_f$ pairs of CW gears for each generation of the SM fermion of flavour $f$, where $f = q,u^c,d^c,l,e^c$. In the FCW mechanism proposed here, the gears of different generations of a given type of SM fermion are allowed to interact with each other maintaining the characteristic nearest neighbour interaction structure of CW mechanism. It is shown that such an arrangement leads to strong or mild intergenerational hierarchies for large or small values of $N_f$. We show that the hierarchies in the masses and mixing of different SM fermions generated by $N_f=4$ for $f=q,u^c,e^c$ and $N_f=2$ for $f=l,d^c$ are in good agreement with the current observations. One can also account for the smallness of $m_b/m_t$ and $m_\tau/m_t$ by introducing appropriate couplings and with the help of FCW mechanism. It is shown that the mechanism provides generalization of the Froggatt-Nielsen mechanism such that the latter with fractional FN charges of fermions can be obtained from the former as an effective theory. The number of gears more or less determines the effective FN charges for all the generations of a given type of fermions.

\begin{acknowledgments} 
I would like to thank Anjan S. Joshipura for useful comments and for a careful reading of the manuscript. This work was supported in part by the SERB Early Career Research Award (ECR/2017/000353) and by a research grant under INSPIRE Faculty Award (DST/INSPIRE/04/2015/000508) from the Department of Science and Technology, Government of India. 
\end{acknowledgments}

\bibliography{references}
\bibliographystyle{apsrev4-1}
\end{document}